\documentclass{article}
\usepackage{arxiv}

\usepackage[utf8]{inputenc} 
\usepackage[T1]{fontenc}    
\usepackage{lmodern}
\usepackage{hyperref}       
\usepackage{url}            
\usepackage{booktabs}       
\usepackage{amsfonts}       
\usepackage{nicefrac}       
\usepackage{microtype}      
\usepackage{lipsum}		
\usepackage{graphicx}
\usepackage[square,sort,comma,numbers]{natbib}
\usepackage{doi}
\usepackage[dvipsnames]{xcolor}%
\usepackage{soul}
\usepackage{subcaption}

\graphicspath{{./figures/}}

\newcommand{\arcade}{{\textbf{\texttt{ARCADE}}}}

\title{\arcade{}: An interactive playground for real-time immersed topology optimization}

\author{ \href{https://orcid.org/0000-0003-2275-6207}{\includegraphics[scale=0.06]{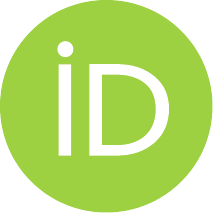}\hspace{1mm}\textcolor{black}{Alejandro M.~Aragón}}\thanks{Corresponding author. E-mail address: \texttt{\href{mailto:a.m.aragon@tudelft.nl}{a.m.aragon@tudelft.nl}} (A.M.~Aragón)}
	\And
	\href{https://orcid.org/0009-0001-5955-8856}{\includegraphics[scale=0.06]{orcid.pdf}\hspace{1mm}\textcolor{black}{Hendrik J.~Algra}} 
	 \AND
	 \\[-1em] 
	 Faculty of Mechanical Engineering, Delft University of Technology\\
	 Mekelweg 2, 2628 CD, Delft, Zuid-Holland,
	 The Netherlands\\	
}

\renewcommand{\shorttitle}{\arcade{}}

\hypersetup{
pdftitle={A template for the arxiv style},
pdfsubject={q-bio.NC, q-bio.QM},
pdfauthor={David S.~Hippocampus, Elias D.~Striatum},
pdfkeywords={First keyword, Second keyword, More},
}

\hypersetup{
  colorlinks=true,
  breaklinks=true,
  urlcolor=RoyalBlue,
  linkcolor=RoyalBlue,
  citecolor=RoyalBlue
} 

\begin{document}
\maketitle

\begin{abstract}

Topology optimization (TO) has found applications across a wide range of disciplines but remains underutilized in practice. Key barriers to broader adoption include the absence of versatile commercial software, the need for specialized expertise, and high computational demands. Additionally, challenges such as ensuring manufacturability, optimizing hyper-parameters, and integrating subjective design elements like aesthetics further hinder its widespread use.

Emerging technologies like augmented reality (AR) and virtual reality (VR) offer transformative potential for TO. By enabling intuitive, gesture-based human-computer interactions, these immersive tools bridge the gap between human intuition and computational processes. They provide the means to integrate subjective human judgment into optimization workflows in real time, creating a paradigm shift toward interactive and immersive design.

Here we introduce the concept of immersive topology optimization (ITO) as a novel design paradigm that leverages AR environments for TO. To demonstrate this concept, we present \arcade{}: Augmented Reality Computational Analysis and Design Environment. Developed in Swift for the Apple Vision Pro mixed reality headset, \arcade{} enables users to define, manipulate, and solve structural optimization problems within an augmented reality setting. By incorporating real-time human interaction and visualization of the design in its intended target location, \arcade{} has the potential to reduce lead times, enhance manufacturability, and improve design integration. Although initially developed for structural optimization, \arcade{}'s framework could be extended to other disciplines, paving the way for a new era of interactive and immersive computational design.

\end{abstract}


\section{Introduction}\label{sec:intro}

Topology optimization (TO) \cite{sigmund2004} has successfully been applied to a myriad of applications spanning many disciplines, including automotive~\cite{yang95}, construction~\cite{amir20}, chip manufacturing~\cite{delissen22}, biomechanical~\cite{koper21}, nanomechanical sensing~\cite{hoj21}, spacecraft~\cite{kim24}, among others. However, the use of TO as a structural design tool is currently more a niche than a standard, as a more widespread adoption in industry is currently hindered by many obstacles. 
One major reason is the lack of TO implementations in commercial software that extend beyond classic compliance minimization problems. Implementing a general topology optimization problem necessitates a high degree of software flexibility, which is currently lacking in software packages. Furthermore, it requires a highly specialized analyst profile with knowledge on computational modeling and optimization. Setting up the problem involves not only implementing the appropriate formulation---i.e., the objective function and its corresponding sensitivity analysis formulation---but also deciding on the type of material interpolation and filtering functions, the definition of appropriate constraints, and the tuning of the optimization hyper-parameters. All these may influence considerably the final design obtained.
Although attempts have been made to alleviate some of these hurdles (e.g., using machine learning~\cite{carstensen24}), achieving a satisfactory optimization result is often as much an art as a science. Partly to blame is also the inherent time-consuming nature the TO design process, which usually requires solving large systems of equations (describing one or multiple physical phenomena) hundreds of times~\cite{Yano:2021aa}. Although there are methods to accelerate this process, for instance by means of surrogate (or meta-) models such as Kriging or machine learning~\cite{Kudela:2022aa},  it is not uncommon to set the design resolution (i.e., the mesh size) to the extent that it strains computers for weeks to obtain the final design.

Another important factor that has received less attention is the trial-and-error process of fine-tuning the final design. Once an optimized design is obtained, challenges may arise during implementation: the design may not be manufacturable (e.g., due to overhang angles), fail to meet certain requirements, or require modifications due to newly available information that needs to be integrated into the design process. While some of these aspects have been addressed, such as incorporating manufacturability constraints into topology optimization formulations~\cite{langelaar16} or by generating a structure composed of predefined structural components~\cite{norato16}, accounting for subjective factors such as aesthetics remains challenging~\cite{Loos:2022aa}. Such aspects are often difficult, if not impossible, to quantify numerically, making it difficult to systematically integrate them into the design process. However, experienced analysts can often quickly assess undesirable characteristics of a particular structure after just a few iterations, highlighting the need for a means to \textit{steer the optimizer} in a different direction as needed, and in real-time. Recent work by Ha and Carstensen \cite{carstensen23} aims in that direction by making a local feature size control available to the designer. In order to take this concept to the next level, and overcome other challenges associated with TO procedures in general, recent advancements in computer technology can be of great assistance.

Over the past decades computers have undergone an extraordinary transformation, both in terms of computing power and their widespread availability. Advances in microchip technology, with an ever-growing number of transistors, have led to an exponential increase in processing capabilities, making devices smaller, faster, and more energy-efficient. Once confined to bulky desktops, computers now fit seamlessly into our daily lives through a wide range of devices, from smart watches and phones, to tablet, laptop, and desktop computers.
The way that people interact with these devices has also changed over time. Alternatives to the traditional mouse and keyboard combination, such as touchscreens and voice commands, are now ubiquitous. There has also been an uptake in enabling users to interact with the digital world by using augmented reality (AR) and virtual reality (VR). Undoubtedly, there is great potential in applying these technologies to topology optimization. From a conceptual perspective, AR and VR integrate the designer's gestures with digital tools, enhancing their ability to work seamlessly with computing systems and therefore closing the gap between human input and machine response.

We envision a future where AR and VR revolutionize design paradigms, enabling a fully immersive and interactive approach that seamlessly integrates human intuition and judgment into the design process in real time. We term this concept immersive topology optimization (ITO), and to bring this vision to life, herein we introduce \arcade{}: Augmented Reality Computational Analysis and Design Environment.

\arcade{} is a computational framework developed in the Swift programming language for the Apple Vision Pro mixed reality headset. It allows users to define, manipulate, and solve structural optimization problems within an immersive AR setting. This interactive environment empowers users to actively engage with the design process, making adjustments in real time. By incorporating human expertise into the optimization workflow and visualizing designs in their intended real-world contexts, \arcade{} has the potential to significantly reduce lead times from conceptualization to manufacturing. While \arcade{} initially deals with structural (compliance) optimization, its versatility could extend to other fields, paving the way for a new era of interactive and immersive computational design.


\section{The \arcade{} application}\label{sec:app}

\arcade{} is our proof-of-concept application that implements immersed topology optimization. Developed for the Apple Vision Pro mixed reality headset, \arcade{} solves a compliance minimization problem in an augmented reality environment. Because the headset is in fact a standalone computer, all processing and visualization is done on the device. The implemented TO formulation follows that of Ferrari and Sigmund~\cite{ferrari20}---in fact we translated their educational code into the Swift programming language. Therefore, we will not delve into the topology optimization formulation but rather focus mostly on the interactive aspects of the application.

Apple Vision Pro uses advanced sensors and cameras to track hand movements in real-time. Therefore, we leverage spatial gesture recognition built into Vision OS for all interactions within the app. A quick thumb-to-index finger pinch is known as a \textit{tap} gesture, and it is used to select or activate an item, similar to a mouse left click. A \textit{pinch-and-drag} gesture is a sustained thumb-to-index finger pinch that is used to grab and move objects or user interface (UI) elements by dragging them through space, akin to a mouse click-and-drag.

\arcade{} consists of a 2-D window with tabs in a navigation menu, where all aspects of the structural compliance problem can be defined, and a 3-D rendering for each corresponding tab. In this section we describe each of the workflow steps, in an order that mirrors the tabs in the app.

\subsection{Defining the domain}

\begin{figure}[!b]
    \centering
    \includegraphics[width=0.85\linewidth]{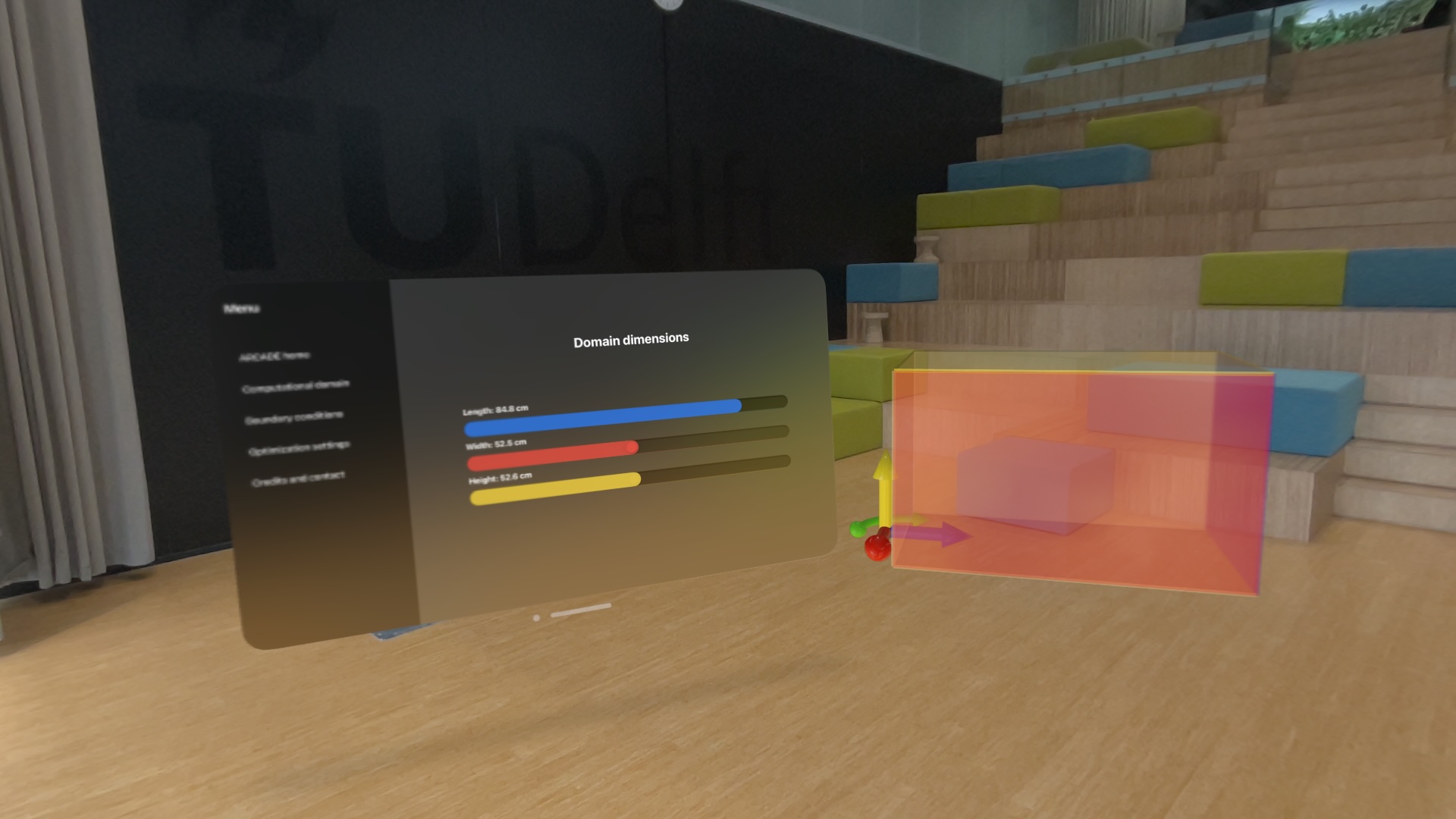}
    \caption{The definition of the design domain takes place in the \texttt{Computational domain} tab. The domain can be resized via the sliders in the window, or simply by pinching and dragging each of the three colored planes of the rendered domain. The three orthogonal arrows represent the Cartesian coordinate system, and pinch-and-drag gestures on the arrows  move the domain in space. The domain can also be rotated around the vertical yellow axis by pinching and dragging the green circular arrow.}
    \label{fig:menu_domain}
\end{figure}

To define the domain we tap the \texttt{Computational domain} menu. The domain, rendered in 3-D as a colored box, appears \textit{floating} in front of the user (see Figure~\ref{fig:menu_domain}). 
Users can resize the domain in one of two ways, either by pinching and dragging the sliders in the 2-D window, or by pinching and dragging the three colored planes of the box (in which the box is resized along a direction perpendicular to that plane). The domain can also be moved in space by pinching and dragging each of the arrows that define the Cartesian coordinate system, or rotated in space (around the yellow axis) by pinching and dragging the circular green arrow.


\subsection{Defining boundary conditions}

\begin{figure}[t]
  \centering
  \begin{subfigure}{0.49\textwidth}
  \centering
    \includegraphics[width=1\linewidth]{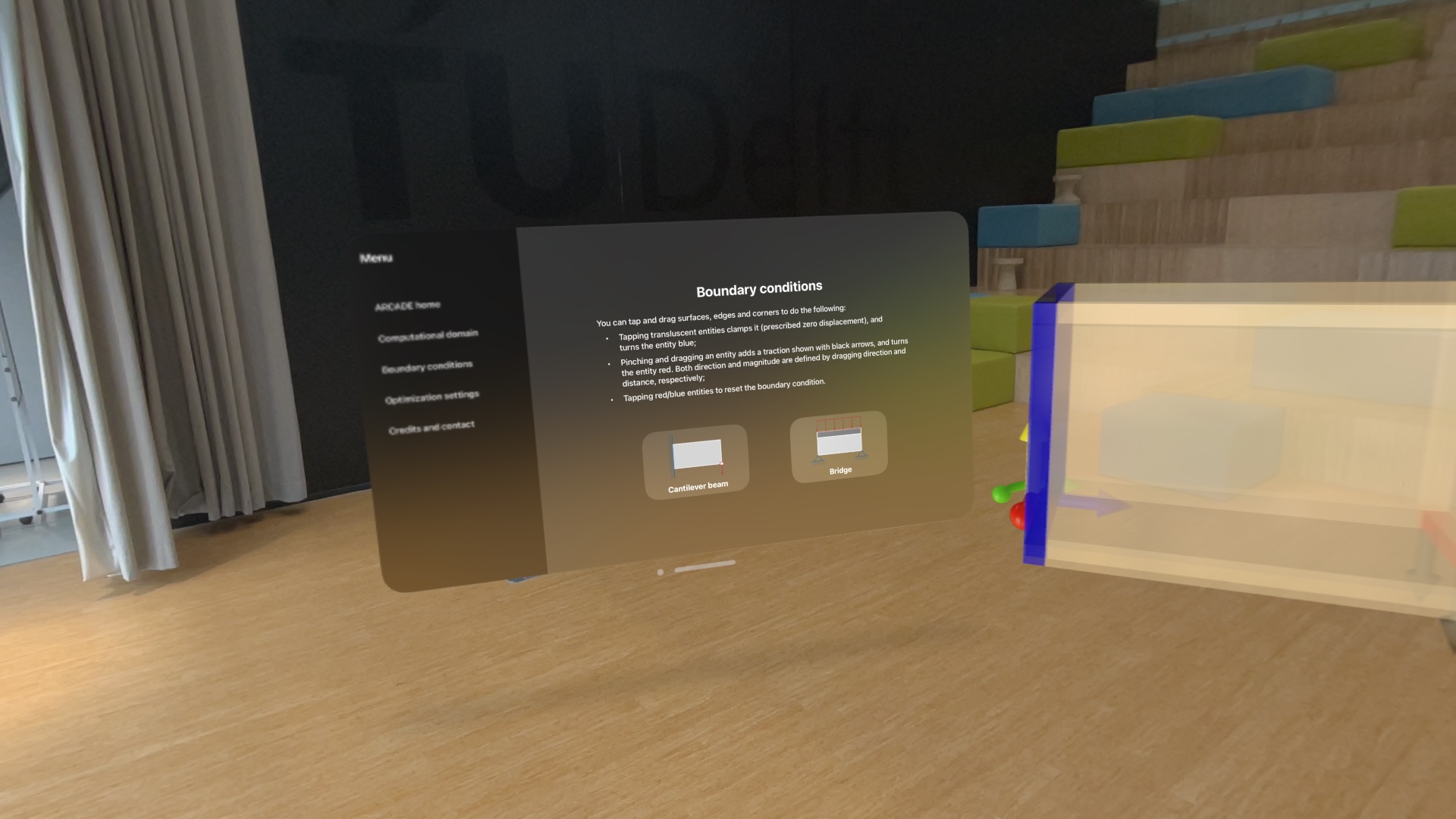}
    \caption{}
    \label{fig:bcs_a}
  \end{subfigure}%
  \hfill
  \begin{subfigure}{0.49\textwidth}
    \centering
    \includegraphics[width=1\linewidth]{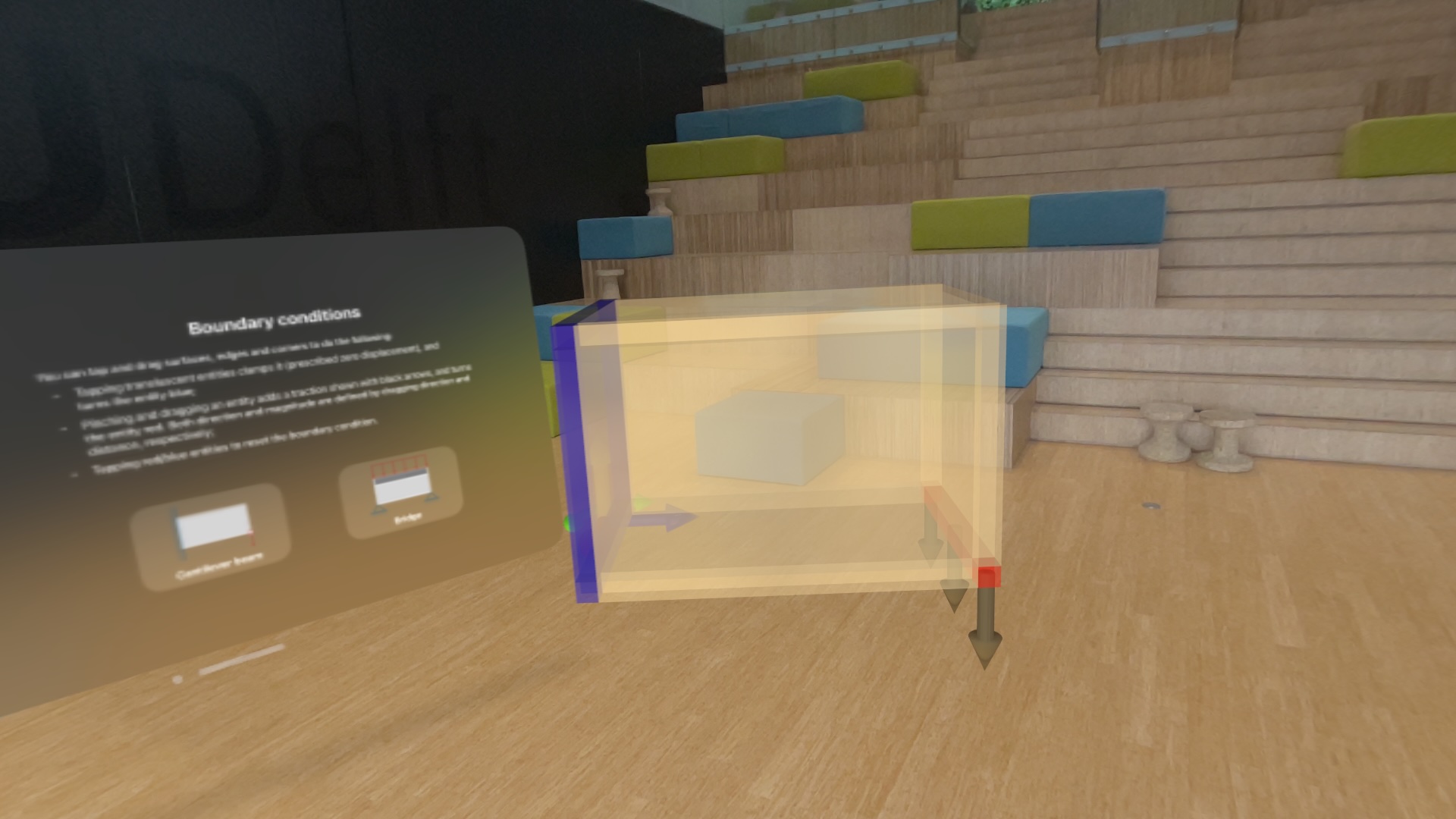}
    \caption{}
    \label{fig:bcs_b}
  \end{subfigure}

  \begin{subfigure}{0.49\textwidth}
  \centering
    \includegraphics[width=1\linewidth]{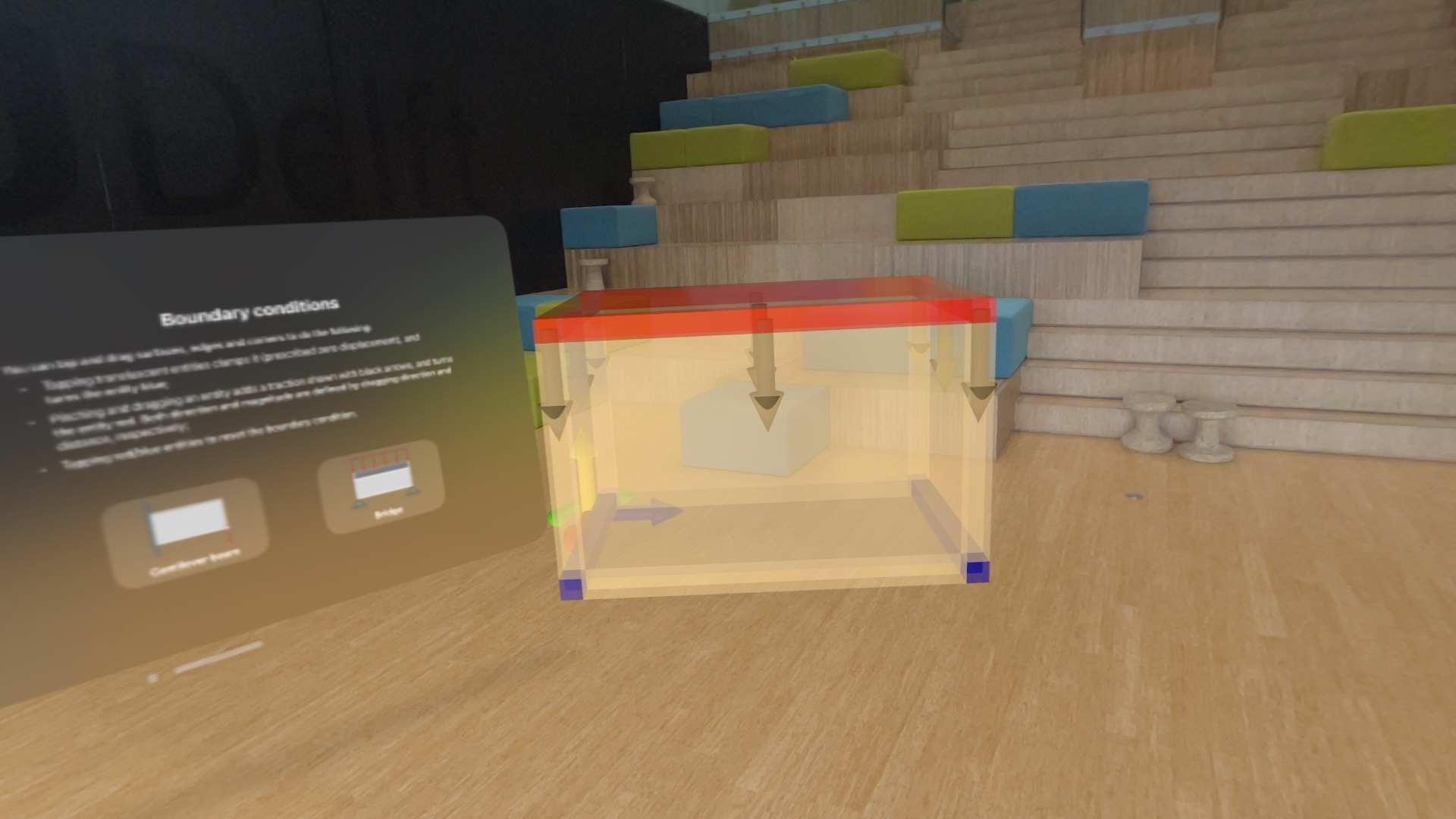}
    \caption{}
    \label{fig:bcs_c}
  \end{subfigure}%
  \hfill
  \begin{subfigure}{0.49\textwidth}
    \centering
    \includegraphics[width=1\linewidth]{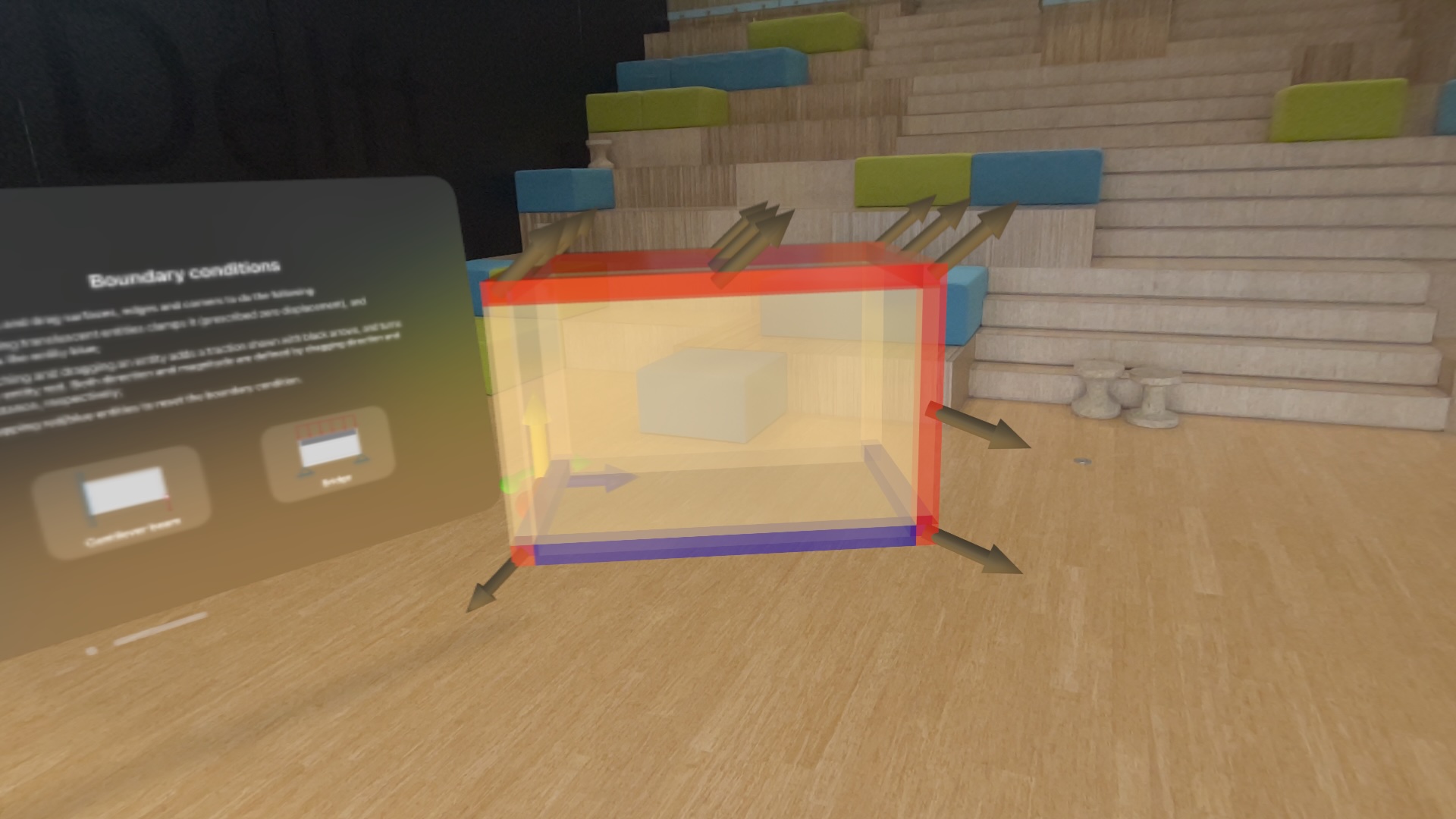}
    \caption{}
    \label{fig:bcs_d}
  \end{subfigure}

\caption{Boundary conditions are defined in the \texttt{Boundary conditions} tab (a). Buttons with default boundary conditions are provided for a cantilever beam (b) and for a bridge (c). Any other boundary conditions can be defined (d): Clamped regions of the boundary, shown in blue, can be defined by simply tapping a vertex, edge, or an entire surface. Boundary tractions (colored in red with black arrows) are defined using pinch-and-drag gestures. Both traction direction and magnitude are defined by the dragging direction and dragging distance, respectively. Tapping a geometric entity with a prescribed boundary condition clears the boundary condition (the entity becomes translucent again).}
\label{fig:bcs}
\end{figure}

Tapping on the \texttt{Boundary conditions} (see Figure~\ref{fig:bcs}) tab allows us to define clamped regions of the boundary (i.e., regions with prescribed zero displacement) and boundary tractions. The rendering of the computational design domain changes to a volumetric representation of faces, edges, and corners, each of which can have a different boundary conditions. The selection is hierarchical, so selecting a surface also selects edges and vertices, and selecting and edge also selects its vertices. Noteworthy, since these two types of boundary conditions are mutually exclusive, only one kind will be active for each geometric entity. Geometric entities without an assigned boundary condition represent regions with zero traction.

The app provides two buttons with default boundary conditions, one for a cantilever beam problem (see Figure~\ref{fig:bcs_b}) and one for a bridge problem (see Figure~\ref{fig:bcs_c}). However, the user can define an arbitrary set of boundary conditions. Simply tapping a geometric entity where a boundary traction is prescribed removes it, and tapping it again clamps it (the entity turns blue). The assigned clamped entity can also be removed simply by tapping again on it. Pinching and dragging an entity defines a traction in the dragged direction and with magnitude proportional to the dragged distance.  Figure~\ref{fig:bcs_d} shows an example where different tractions are prescribed on the top surface, on of the right edges, and the bottom-left vertex. 

\subsection{Defining and running the optimization}

\begin{figure}[t]
    \centering
    \includegraphics[width=0.85\linewidth]{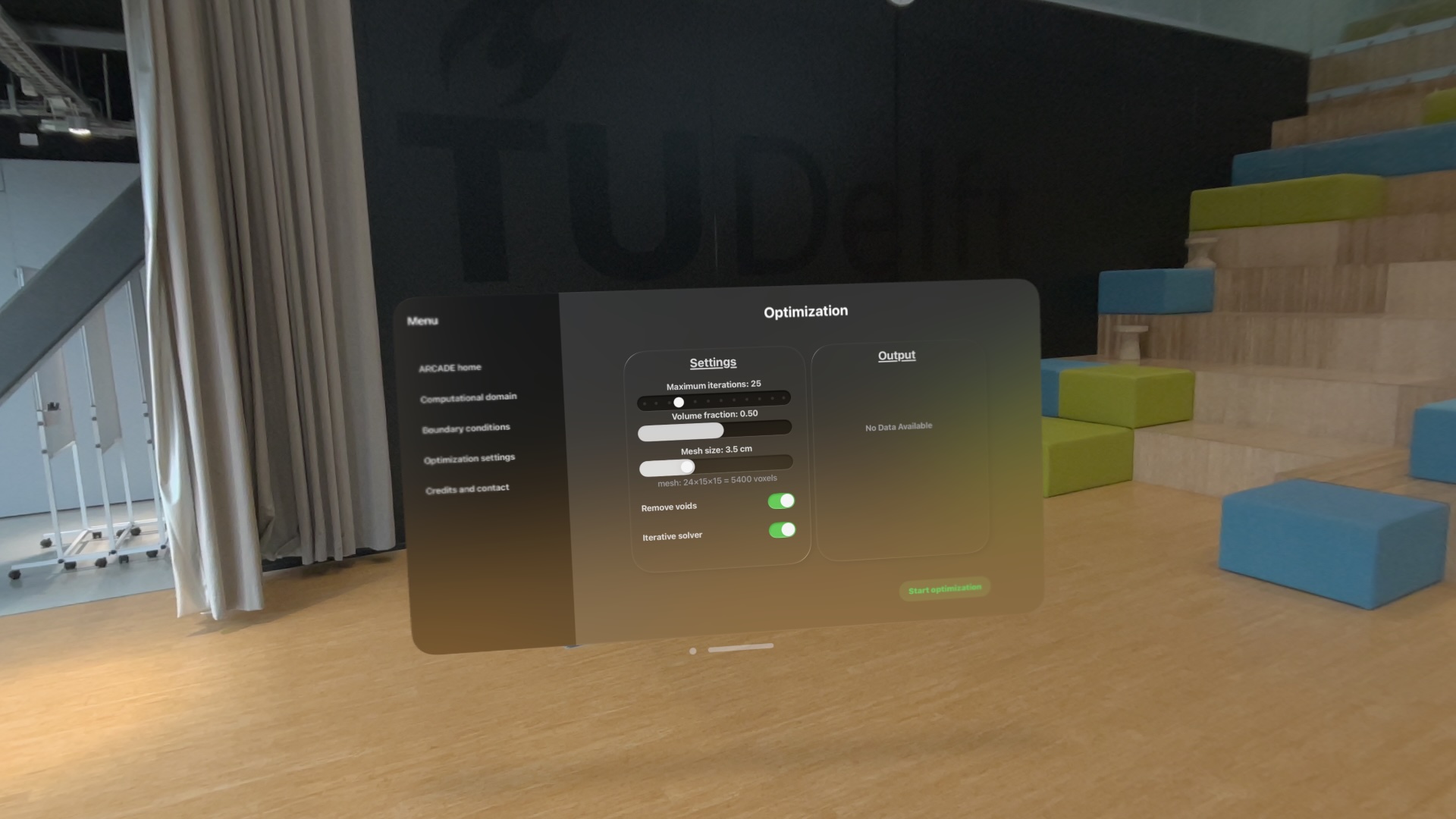}
    \caption{Settings that define the optimization are set in the \texttt{Optimization settings} tab. These include the maximum allowed number of iterations, the target volume fraction of material within the computational design domain, and the finite element mesh size used for the discretization. The toggles allows the removal of voids during optimization and to use an iterative solver (both set to true by default).}
    \label{fig:menu_optimize}
\end{figure}

After defining the computational domain dimensions and location in space, and the boundary conditions, we tap on the \texttt{Optimization settings} tab (see Figure \ref{fig:menu_optimize}). Settings include the maximum number of iterations, the target volume fraction (as a percentage of the computational domain), and the finite element mesh size. These can be changed by pinching and dragging the sliders. Changing the element size updates the number of voxels in the finite element discretization. Two toggle buttons add the ability to remove voids during the optimization and to use an iterative solver instead of a direct solver (both toggles are set to true by default).

\begin{figure}[!b]
  \centering
  \begin{subfigure}{0.49\textwidth}
  \centering
    \includegraphics[width=1\linewidth]{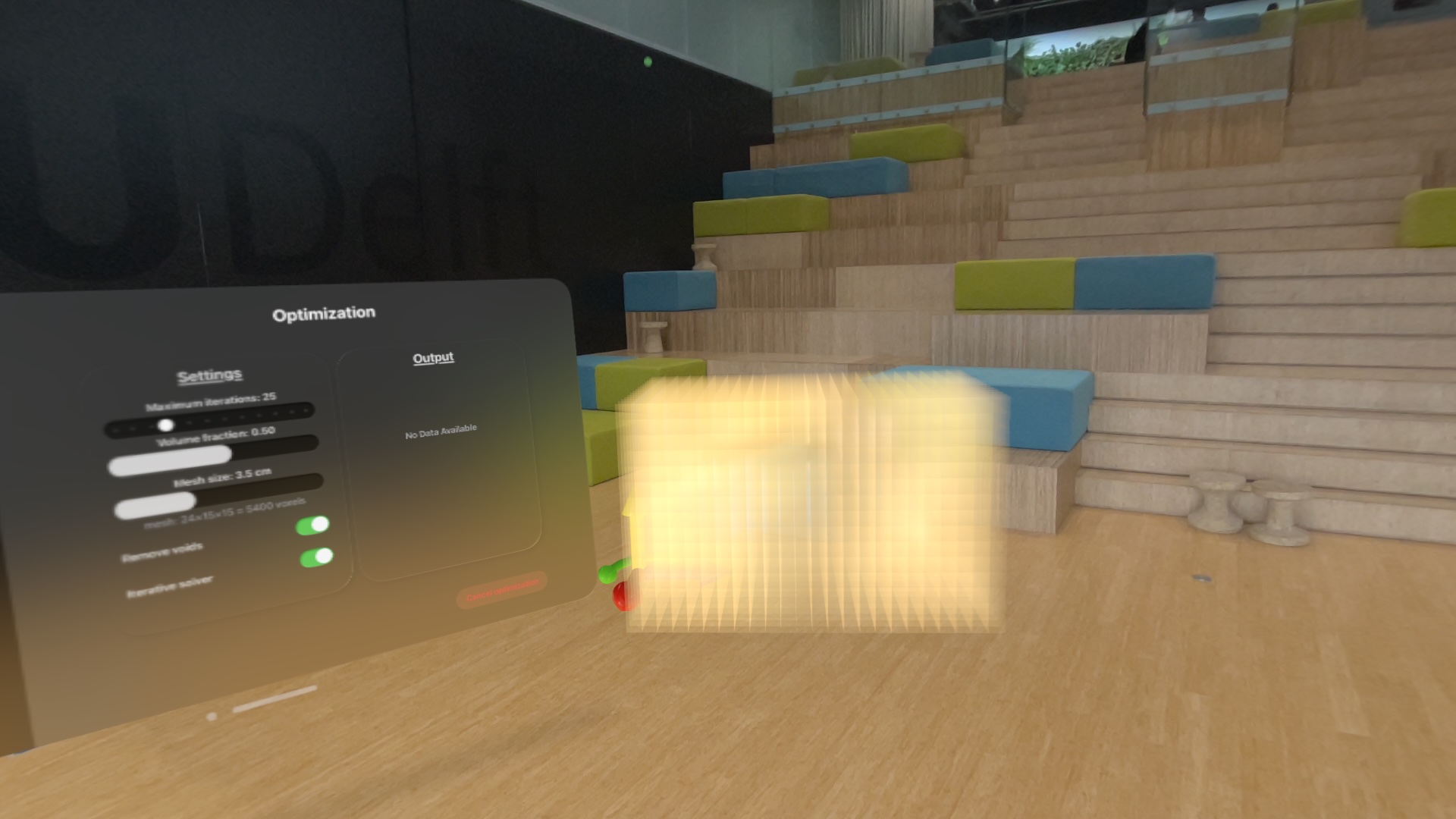}
    \caption{}
    \label{fig:optimization_a}
  \end{subfigure}%
  \hfill
  \begin{subfigure}{0.49\textwidth}
    \centering
    \includegraphics[width=1\linewidth]{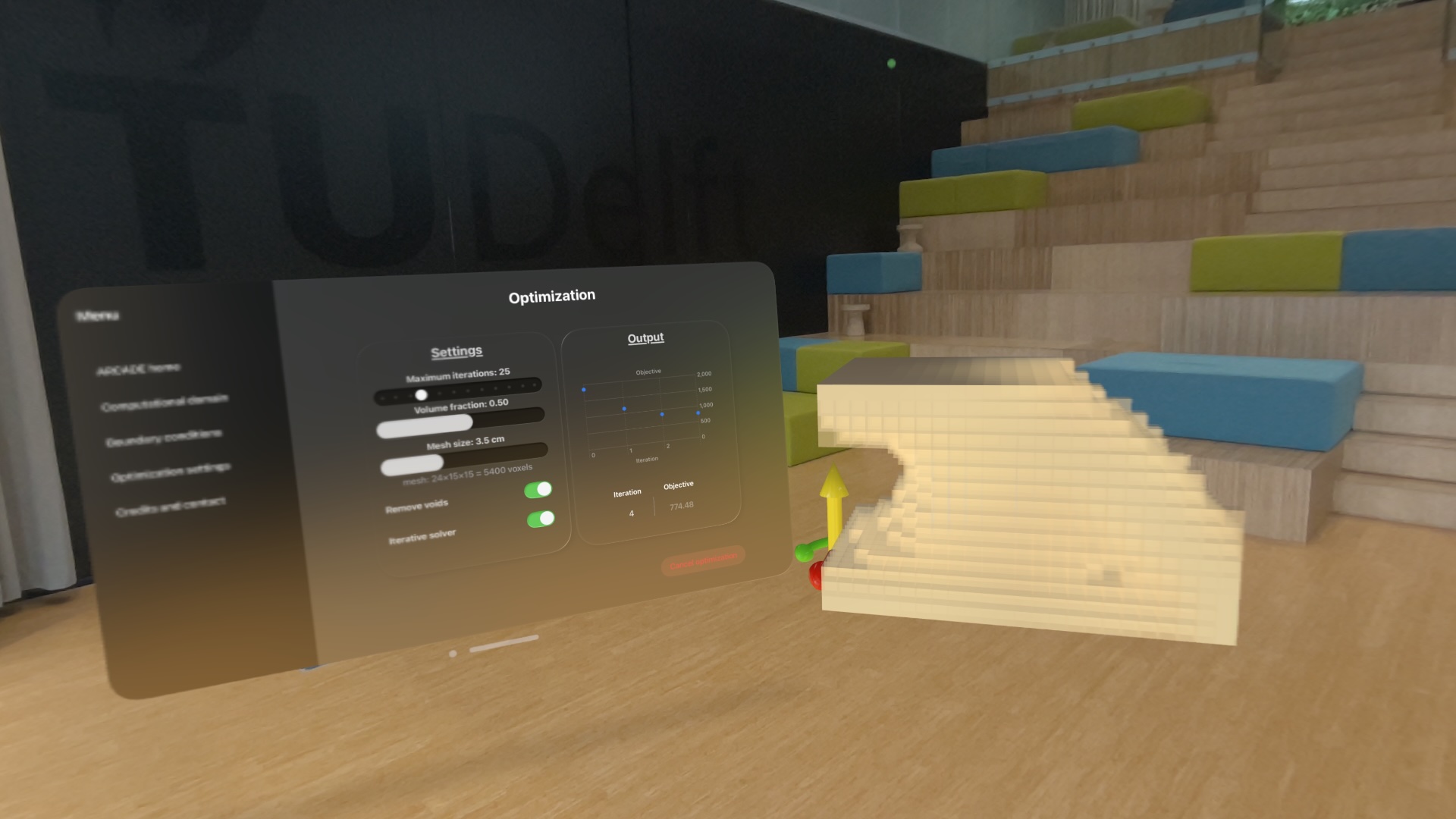}
    \caption{}
    \label{fig:optimization_b}
  \end{subfigure}

  \begin{subfigure}{0.49\textwidth}
  \centering
    \includegraphics[width=1\linewidth]{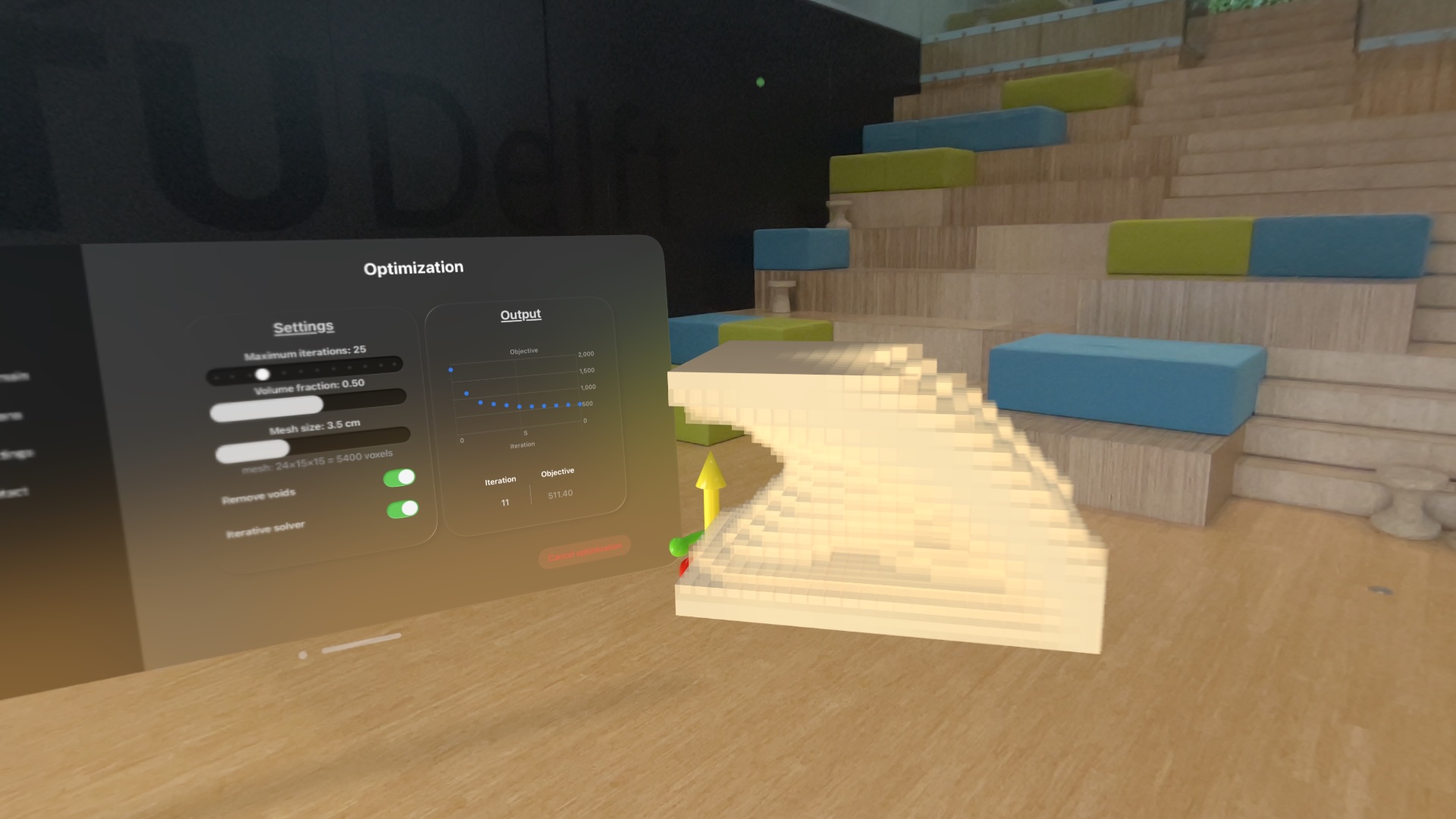}
    \caption{}
    \label{fig:optimization_c}
  \end{subfigure}%
  \hfill
  \begin{subfigure}{0.49\textwidth}
    \centering
    \includegraphics[width=1\linewidth]{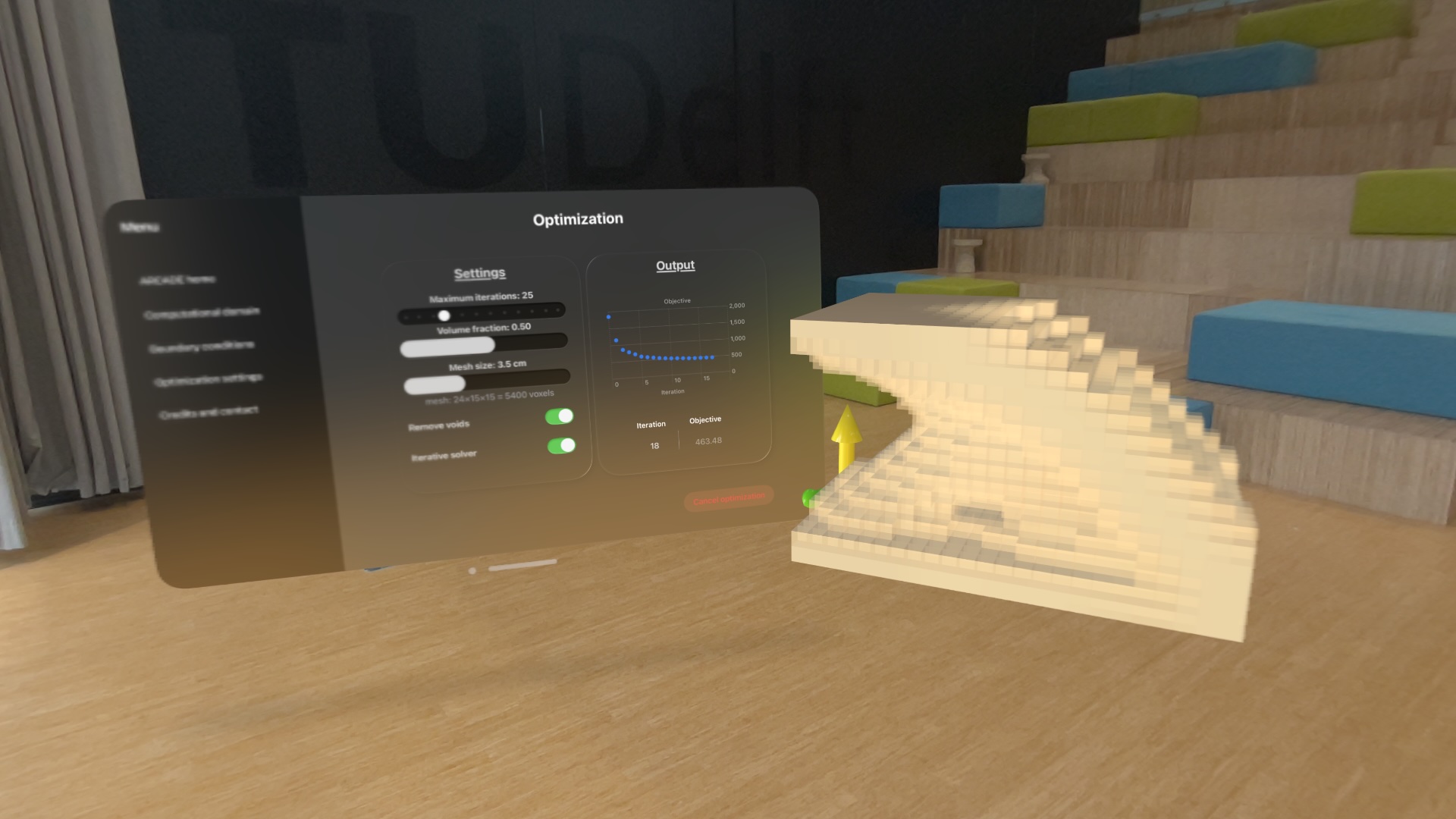}
    \caption{}
    \label{fig:optimization_d}
  \end{subfigure}

\caption{During the optimization process, which starts by pressing the \texttt{Start optimization} button, changes to the design are visualized in real-time. The objective function value as a function of iteration is displayed in the floating 2-D menu. The app is interactive so that changes to boundary conditions can take place during the optimization.}
\label{fig:optimization}
\end{figure}

When satisfied with the settings, the user taps the button to start the immersed topology optimization. 
Throughout the optimization process, changes in the design are rendered in real time and the corresponding objective function value is shown in the \texttt{Output} section is shown as a function of iteration.
The app is interactive, so during the optimization the user can go back to the \texttt{Boundary conditions} tab so that changes take effect immediately. Behind the scenes, changing boundary tractions simply modifies the force vector---i.e., the right-hand side of the system of linear equations that is solved. Changing clamped regions takes more effort as it necessitates finding the new set of degrees of freedom whose displacement is prescribed to zero.
We also built sounds for each optimization step (Rubik's cube sound) and for the completion of the optimization. Figure~\ref{fig:optimization} illustrates the optimization of the cantilever beam problem.

\begin{figure}[t]
  \centering
  \begin{subfigure}{0.49\textwidth}
  \centering
    \includegraphics[width=1\linewidth]{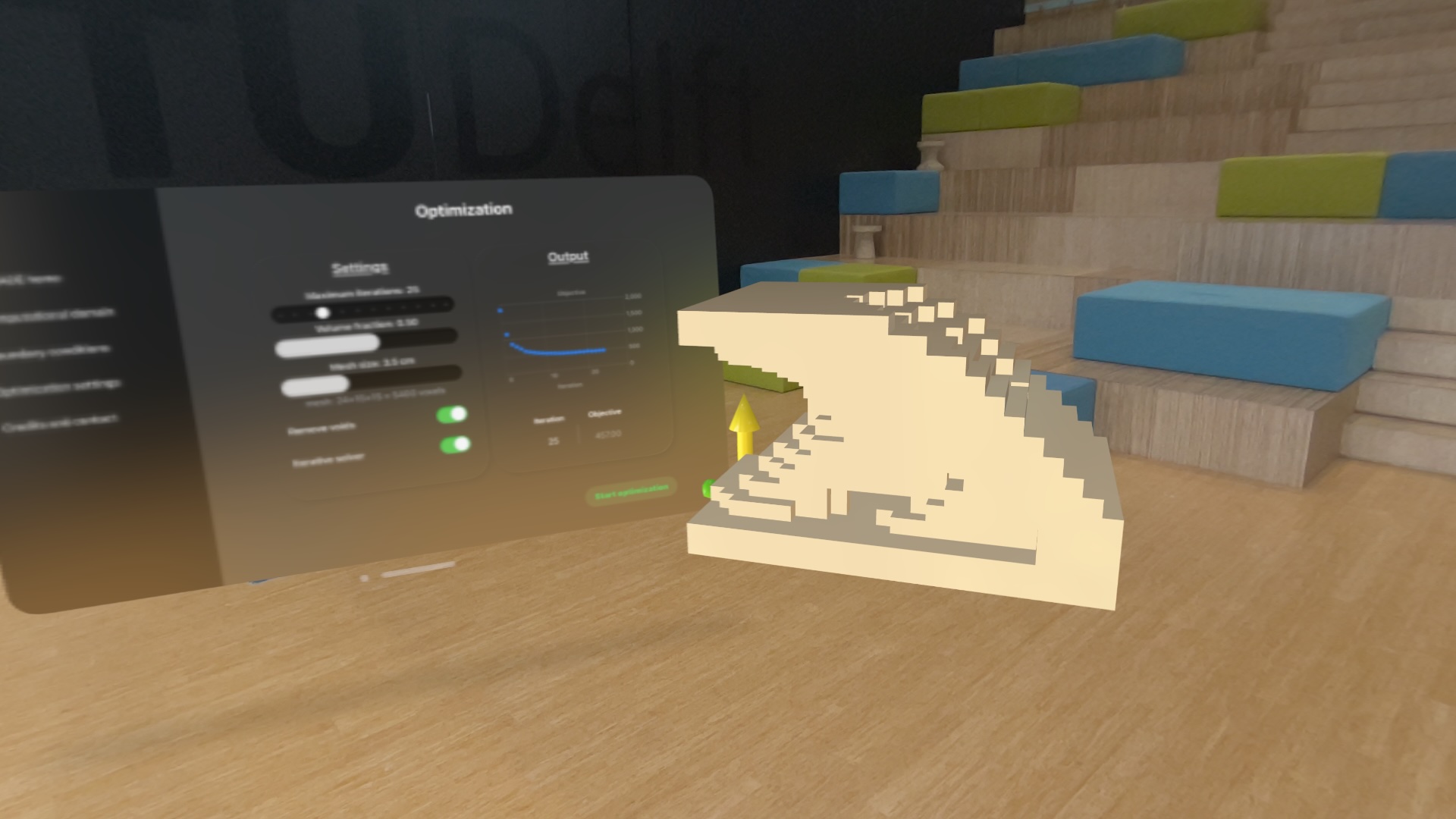}
    \caption{}
    \label{fig:final_design_a}
  \end{subfigure}%
  \hfill
  \begin{subfigure}{0.49\textwidth}
    \centering
    \includegraphics[width=1\linewidth]{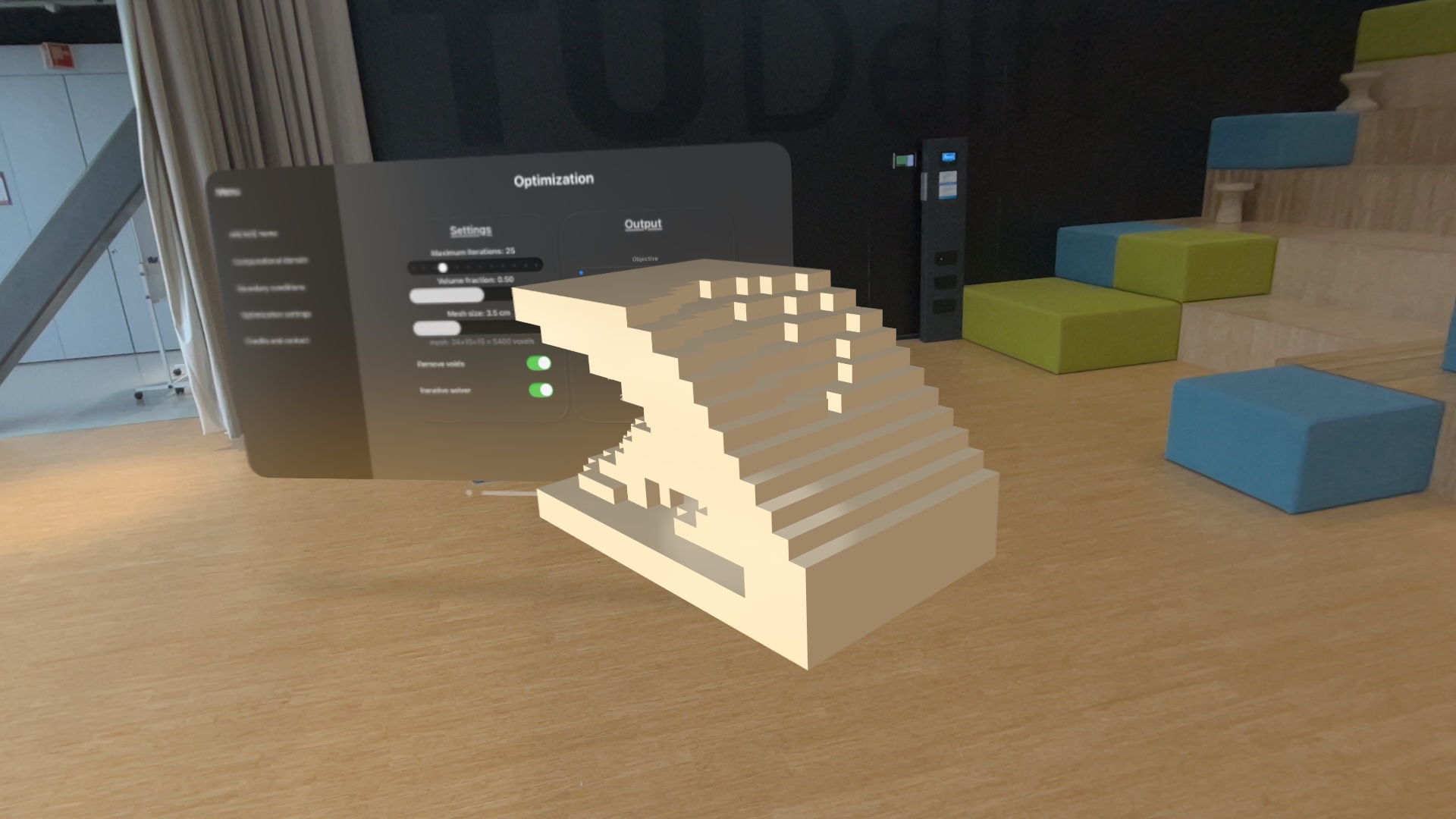}
    \caption{}
    \label{fig:final_design_b}
  \end{subfigure}

  \begin{subfigure}{0.49\textwidth}
  \centering
    \includegraphics[width=1\linewidth]{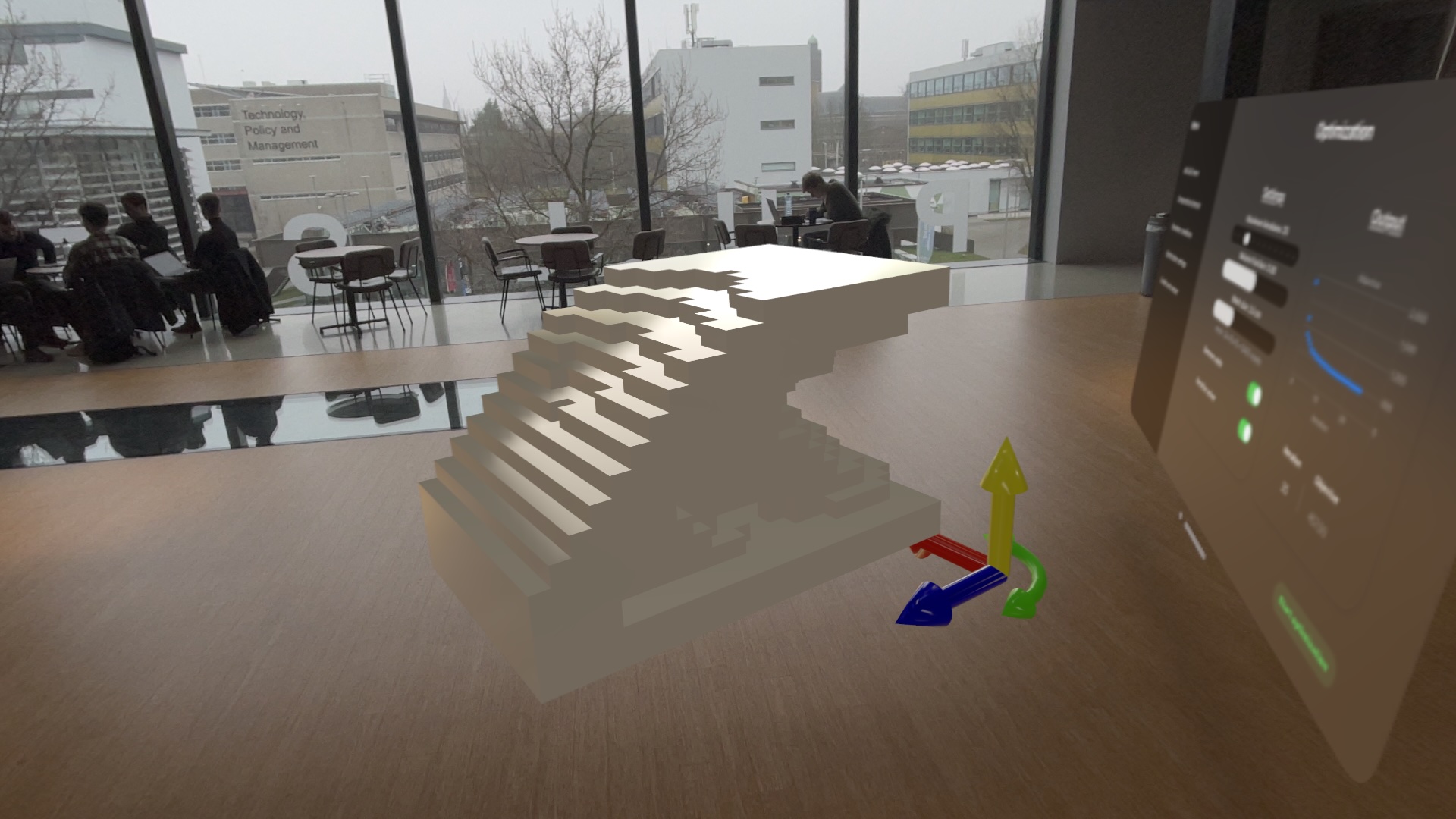}
    \caption{}
    \label{fig:final_design_c}
  \end{subfigure}%
  \hfill
  \begin{subfigure}{0.49\textwidth}
    \centering
    \includegraphics[width=1\linewidth]{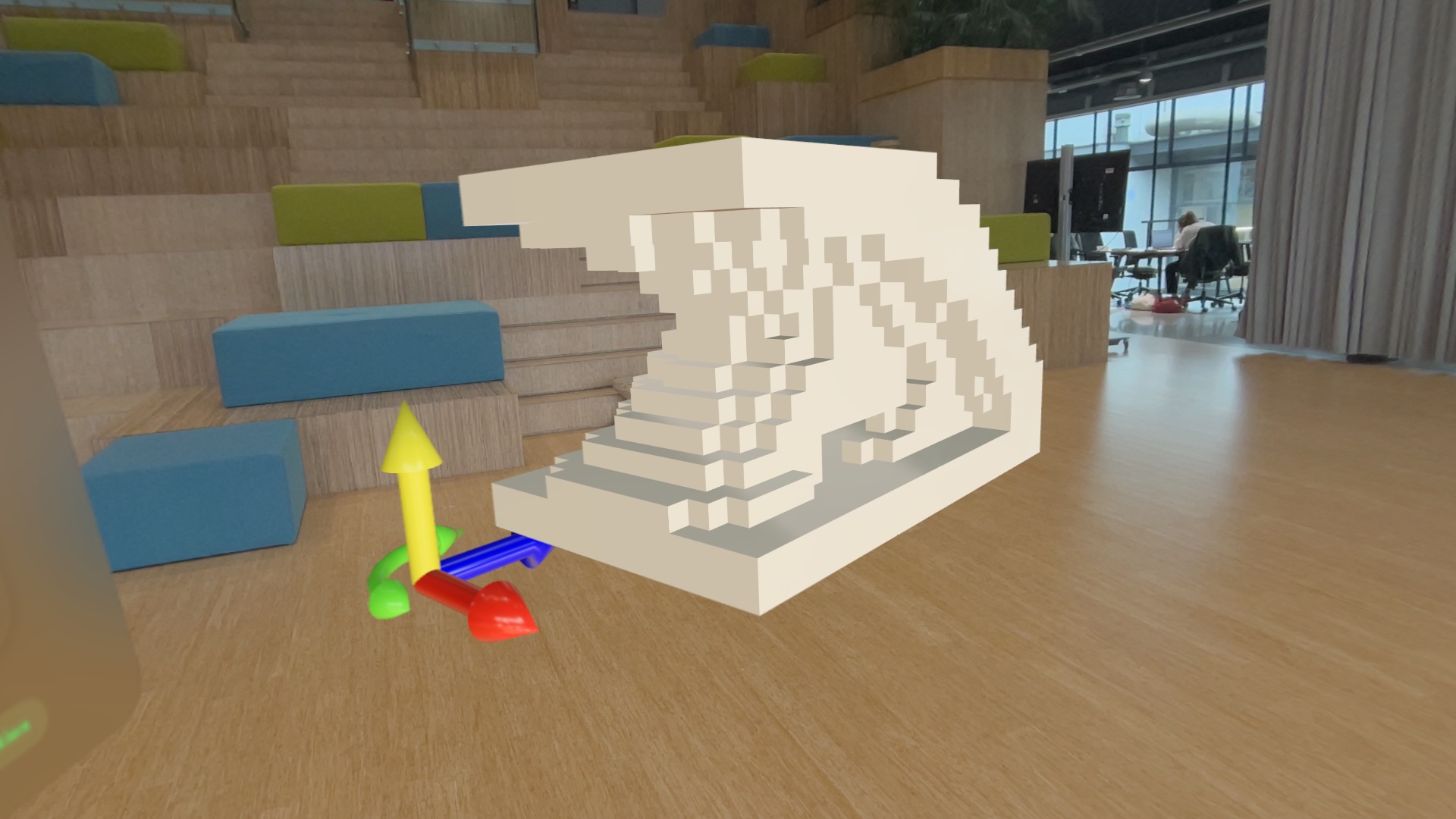}
    \caption{}
    \label{fig:final_design_d}
  \end{subfigure}

\caption{The final result of an optimization procedure fully renders all elements with a density above a pre-defined threshold. The design stays still in space so it can be inspected from any angle.}
\label{fig:final_design}
\end{figure}

At the end of the optimization, the final designs is rendered as a fully \textit{black-and-white} design (see Figure~\ref{fig:final_design}). Notice that the final design remains fixed in space so that we can walk around to inspect it from any angle.

\subsection{Topology optimization formulation and implementation}
As mentioned earlier, our topology optimization code is essentially a translated and modified version of the educational 3-D TO code by Ferrari and Sigmund~\cite{ferrari20}. As the original code relied on built-in \texttt{MATLAB} functions that are not available in Swift (such as \texttt{imfilter} for the filtering procedure), similar functionalities were developed using Apple's framework. In the case of \texttt{imfilter}---used in the original MATLAB code to perform a 3-D filtering operation on the vector of design densities---we used Apple's \texttt{Accelerate} framework to perform a similar filtering procedure. In our implementation, we perform 3-D filtering as a sequence of 2-D convolutions. We note that, while we opted for zero-padding at boundaries, the MATLAB \texttt{imfilter} function also supports other types of padding. Since Swift has a lower level of abstraction, the translation significantly increased the number of coding lines beyond the reference 125 lines. Additionally, a different node numbering convention was adopted, which greatly simplified the identification of unique faces, edges and vertices, and the subsequent definition of boundary conditions.

\section{Swift implementation details}

The goal of this section is to provide insight into \arcade{}'s application structure.
The operating system of the Apple Vision Pro is VisionOS~\cite{apple_visionos}, so the technical terms used in this chapter often refer to concepts and components specific to its application developer environment.

\subsection{User interface}
The floating menu that appears when opening the application is generated using a \texttt{Window} from the \texttt{SwiftUI} framework. The window contains a \texttt{NavigationSplitView} that defines the different tabs of the menu. Switching between tabs loads new content, and asynchronously instructs the \texttt{DomainManager} (discussed below) to load or unload its 3-D content.

\subsection{Rendering 3-D content}
In visionOS, unbounded content in a user’s surroundings---such as the design domain in our application---is managed within an \texttt{ImmersiveSpace}~\cite{apple2024immersive}. 
This space acts as an invisible coordinate system that allows for the dynamic addition or removal of 3-D content.
Upon launching the application, an \texttt{ImmersiveSpace} is created with its local coordinate system centered on the user’s initial position. The \texttt{ImmersiveSpace} remains active throughout the session, allowing the user to move freely while ensuring that rendered 3D content remains fixed in its designated position within the space.

In order to load and unload 3-D content, an empty \texttt{Entity} from the \texttt{RealityKit} framework is added to the \texttt{ImmersiveSpace} during its initialization. During the flow of the application, all 3-D content is added or removed by making it a child of this \texttt{Entity}.

The logic behind loading and unloading different domain renderings is taken care of by the \texttt{DomainManager} class. The 3-D content itself is produced by three other classes, namely the \texttt{ResizableDomain} (for reshaping the domain in the \texttt{Computational domain} tab), \texttt{BCDomain} (for setting boundary conditions in the \texttt{Boundary conditions} tab), and \texttt{MeshDomain} (for rendering the finite element mesh during optimization).  Each of these classes keep the state (properties in Swift) related to the content that they render. For example, the \texttt{ResizableDomain} object stores the actual dimensions of the domain, and the \texttt{MeshDomain} the finite element mesh size. Furthermore, the gestures that allow the user to interact with the 3-D content are also defined in these classes.

\subsection{Running the optimization}
Everything related to the finite element analysis and optimization is managed by the \texttt{SimulationManager} class. When the user starts the simulation from the 2-D menu, this class starts running the iterative topology optimization code on a separate thread. This allows the user to keep using the application while the optimization is in progress. A very useful property of the \texttt{SwiftUI} framework is that objects can become \texttt{Observable}, which enables a \textit{reactive programming model} where UI components automatically update when the underlying data of \texttt{Observable} objectes change.
For example, as the topology optimization progresses, the density vector variable stored by the \texttt{SimulationManager} changes. This prompts the \texttt{DomainManager.MeshDomain} to update the 3-D rendering of the mesh. Furthermore, changes made in the optimization tab of the 2-D menu directly change the stored variables of \texttt{SimulationManager}. This means that changes made to the volume constraint or iteration limit can directly be taken into account as the optimization progresses in real-time.

\subsection{App availability}
We have made \arcade{} freely available on the visionOS App Store. As the writing of this article, \arcade{} can run in visionOS versions 2.0-2.2.


\section{Discussion}\label{sec:discussion}
Immersed topology optimization in an augmented reality environment has the potential to shift design paradigms.
By incorporating subjective human aspects early into the design process, visualizing the design in its intended target location, and by enabling changes to the optimization process in real-time, this technology offers the potential to accelerate the design process to reach a final prototype. \arcade{} represents our endeavor to make a contribution in that direction.

By allowing users to use hand gestures to set up the domain, prescribe boundary conditions, and solve a topology optimization problem, \arcade{} reduces the steep learning curve associated with traditional modeling tools. This feature may in turn make topology optimization more accessible to users that do not particularly have a strong background in mechanics, e.g., interior designers, allowing them to focus on other aspects such as aesthetics.

While in our first implementation it is possible to make changes to boundary conditions in real time, we plan to make other aspects of the design process interactive. For instance, we plan to change the size of the domain in real time, which would change the entire finite element mesh structure and therefore the corresponding structure of the linear system of equations that is solved behind the scenes.
It is also our hope that \arcade{} serves as a source of inspiration for other developments.
For instance, the augmented reality setting in which the optimization takes place can elevate existing concepts that integrate human knowledge into optimization processes, such as setting the size of local features~\cite{carstensen23}, to the next level.

All in all, the enhanced accessibility, expedited design process, and improved decision-making capabilities that arise from a seamless user interaction within an immersive design environment hold significant potential.
We hope that this innovative design paradigm be embraced by the optimization community, facilitating the transition of our current research endeavors from an educational article to a valuable tool in the design toolkit of researchers and industry professionals.

\bibliographystyle{unsrt}
\bibliography{references}

\end{document}